\documentclass[aip,reprint]{revtex4-1} 
\usepackage{rotate}
\usepackage{rotating}
\usepackage{graphicx}
\usepackage{wrapfig}
\usepackage{comment}
\usepackage[caption=false]{subfig}
\usepackage{longtable}
\usepackage{color}
\usepackage{appendix}
\usepackage{setspace}
\usepackage{array}
\usepackage{siunitx}
\usepackage[sc]{mathpazo}
\usepackage[T1]{fontenc}

\DeclareSIUnit\po{$\times$10$^{-13}$}
\DeclareSIUnit\microncubo{$\mu m^3$}
\newcommand{\phase}{$\delta\phi$}
\DeclareSIUnit\nm{nm}
\DeclareSIUnit\eV{eV}

\newcommand{\cobalt}{$^{57}$Co}
\newcommand{\musec}{$\mu s$}

\newcommand{\ka}{KID-1}
\newcommand{\kb}{KID-2}
\newcommand{\kc}{KID-3}
\newcommand{\kd}{KID-4}
\newcommand{\calder}{CALDER}
\newcommand{\tauqp}{$\tau_{qp}$}
\newcommand{\qp}{quasiparticles}

\newcommand{\FilmThickness}{\SI{40}{\nm}}
\newcommand{\SubstrateThickness}{300\,$\mu$m}

\newcommand{\Nzero}{1.72$\times$10$^{10}$\,eV$^{-1}\mu m^{-3}$}
\newcommand{\NoiseGlobalResolution}{154$\pm7$\,eV}
 
\newcommand{\efficiency}{(18$\pm$2)\,$\%$}

\begin{document}

\title{Energy resolution and efficiency of phonon-mediated  KIDs for light detection}
\author{L.~Cardani}
\email{laura.cardani@roma1.infn.it}
\affiliation{Dipartimento di Fisica - Sapienza Universit\`{a} di Roma, Piazzale Aldo Moro 2, 00185, Roma - Italy}
\affiliation{Physics Department - Princeton University, Washington Road, 08544, Princeton - NJ, USA}
\author{I.~Colantoni}
\affiliation{Dipartimento di Fisica - Sapienza Universit\`{a} di Roma, Piazzale Aldo Moro 2, 00185, Roma - Italy}
\author{A.~Cruciani}
\affiliation{Dipartimento di Fisica - Sapienza Universit\`{a} di Roma, Piazzale Aldo Moro 2, 00185, Roma - Italy}
\affiliation{INFN - Sezione di Roma, Piazzale Aldo Moro 2, 00185, Roma - Italy}
\author{S.~Di Domizio}
\affiliation{Dipartimento di Fisica - Universit\`{a} degli Studi di Genova, Via Dodecaneso 33, 16146, Genova - Italy}
\affiliation{INFN - Sezione di Genova, Via Dodecaneso 33, 16146, Genova - Italy}
\author{M.~Vignati}
\affiliation{Dipartimento di Fisica - Sapienza Universit\`{a} di Roma, Piazzale Aldo Moro 2, 00185, Roma - Italy}
\affiliation{INFN - Sezione di Roma, Piazzale Aldo Moro 2, 00185, Roma - Italy}
\author{F.~Bellini}
\affiliation{Dipartimento di Fisica - Sapienza Universit\`{a} di Roma, Piazzale Aldo Moro 2, 00185, Roma - Italy}
\affiliation{INFN - Sezione di Roma, Piazzale Aldo Moro 2, 00185, Roma - Italy}
\author{N.~Casali}
\affiliation{Dipartimento di Fisica - Sapienza Universit\`{a} di Roma, Piazzale Aldo Moro 2, 00185, Roma - Italy}
\affiliation{INFN - Sezione di Roma, Piazzale Aldo Moro 2, 00185, Roma - Italy}
\author{M.G.~Castellano}
\affiliation{Istituto di Fotonica e Nanotecnologie - CNR, Via Cineto Romano 42, 00156, Roma - Italy}
\author{A.~Coppolecchia}
\affiliation{Dipartimento di Fisica - Sapienza Universit\`{a} di Roma, Piazzale Aldo Moro 2, 00185, Roma - Italy}
\author{C.~Cosmelli}
\affiliation{Dipartimento di Fisica - Sapienza Universit\`{a} di Roma, Piazzale Aldo Moro 2, 00185, Roma - Italy}
\affiliation{INFN - Sezione di Roma, Piazzale Aldo Moro 2, 00185, Roma - Italy}
\author{C.~Tomei}
\affiliation{INFN - Sezione di Roma, Piazzale Aldo Moro 2, 00185, Roma - Italy}

\begin{abstract}
The development of sensitive cryogenic light detectors is of primary interest for bolometric experiments searching for rare events like dark matter interactions or neutrino-less double beta decay.
Thanks to their good energy resolution and the natural multiplexed read-out, Kinetic Inductance Detectors (KIDs) are particularly suitable for this purpose.
To efficiently couple KIDs-based light detectors to the large crystals used by the most advanced bolometric detectors, active surfaces of several cm$^2$ are needed. 
For this reason, we are developing phonon-mediated detectors.
In this paper we present the results obtained with a prototype consisting of four \FilmThickness\ thick aluminum resonators patterned on a 2$\times$2\,cm$^2$ silicon chip, and
calibrated with optical pulses and  X-rays. The detector features a noise resolution $\sigma_E=154\pm7$\,eV\ and an \efficiency\ efficiency.
\end{abstract}

\maketitle

\section{Introduction}
\label{sec:introduction}
From their first applications in photon detection\cite{Day:2003fk}, 
Kinetic Inductance Detectors (KIDs) became rapidly subject of several R$\&$D activities in different physics sectors such as astrophysics\cite{Monfardini2011,Mazin:2013wvi} and search of dark matter interactions\cite{Golwala2008}, proving to be very versatile devices\cite{zmu_annrev2012}.
In superconductors biased with an AC current, Cooper pairs oscillate and acquire kinetic inductance.
Interactions with energy larger than the binding energy of Cooper pairs (2$\Delta_0$) can break them into quasiparticles, producing a variation in the kinetic inductance.
The superconductor is inserted in a high merit factor LC circuit, and the energy of the interactions can be reconstructed by monitoring the variations in the resonance parameters.

The key features of KIDs consist in the good intrinsic energy resolution, and in the possibility of easily tuning each detector to a different resonant frequency. 
This natural aptitude to frequency multiplexed read-out allows operation of a few hundreds of KIDs with two cables and one cryogenic amplifier. 
In addition, since the electronics is located at room temperature, the installation of KIDs arrays would require minor modifications to the pre-existing cryogenic facilities.
Another important advantage of these devices is that they are operated well below the superconductor critical temperature, where the quasiparticle lifetime and the internal quality factor saturate, resulting in a negligible dependence on the temperature.

Light detectors (LDs) based on KIDs could enable particle identification, and therefore background suppression\cite{CUPIDRD,CUPIDsci} in bolometric experiments searching for neutrino-less double beta decay (0$\nu\beta\beta$) or dark matter interactions, such as CUORE\cite{Artusa:2014lgv} and LUCIFER\cite{Beeman:2013sba}. 
 
The present limit of KIDs in this kind of application resides in their small active surface (few mm$^2$).
Macro-bolometers used by CUORE and LUCIFER feature a surface of several cm$^2$,
thus an effective read-out of the light emitted by these crystals demands for LDs with large active area.
Obtaining a similar surface with KIDs would require hundreds of pixels for each LD, which is not a realistic option for experiments with $\sim$1000 detectors.
This problem can be overcome following the phonon-mediated approach developed for Neutron Tansmutation Doped Ge thermistors\cite{Fiorini1984} and for Transition Edge Sensors\cite{Proebst1995}, and recently proposed also for KIDs\cite{swenson,moore2}.

The \calder\ project\cite{CalderWhitePaper} aims to demonstrate that such an approach allows to develop large-area LD with noise energy resolution $<$20\,eV RMS, multiplexed read-out, high reproducibility and scalability.

In this paper we present the results of a \calder\ development prototype, implementing four aluminum resonators. We perform an absolute calibration of the energy absorbed by the KIDs and derive the efficiency of the detector.
We characterize the response with photons at 400\,nm and X-rays from a \cobalt\ source. Finally, we analyze and evaluate the energy resolution of the detector.

\section{Detector Fabrication}
\label{sec:Detector}
\begin{figure}[htbp]
 \includegraphics[width=.45\textwidth, natwidth=842, natheight=595]{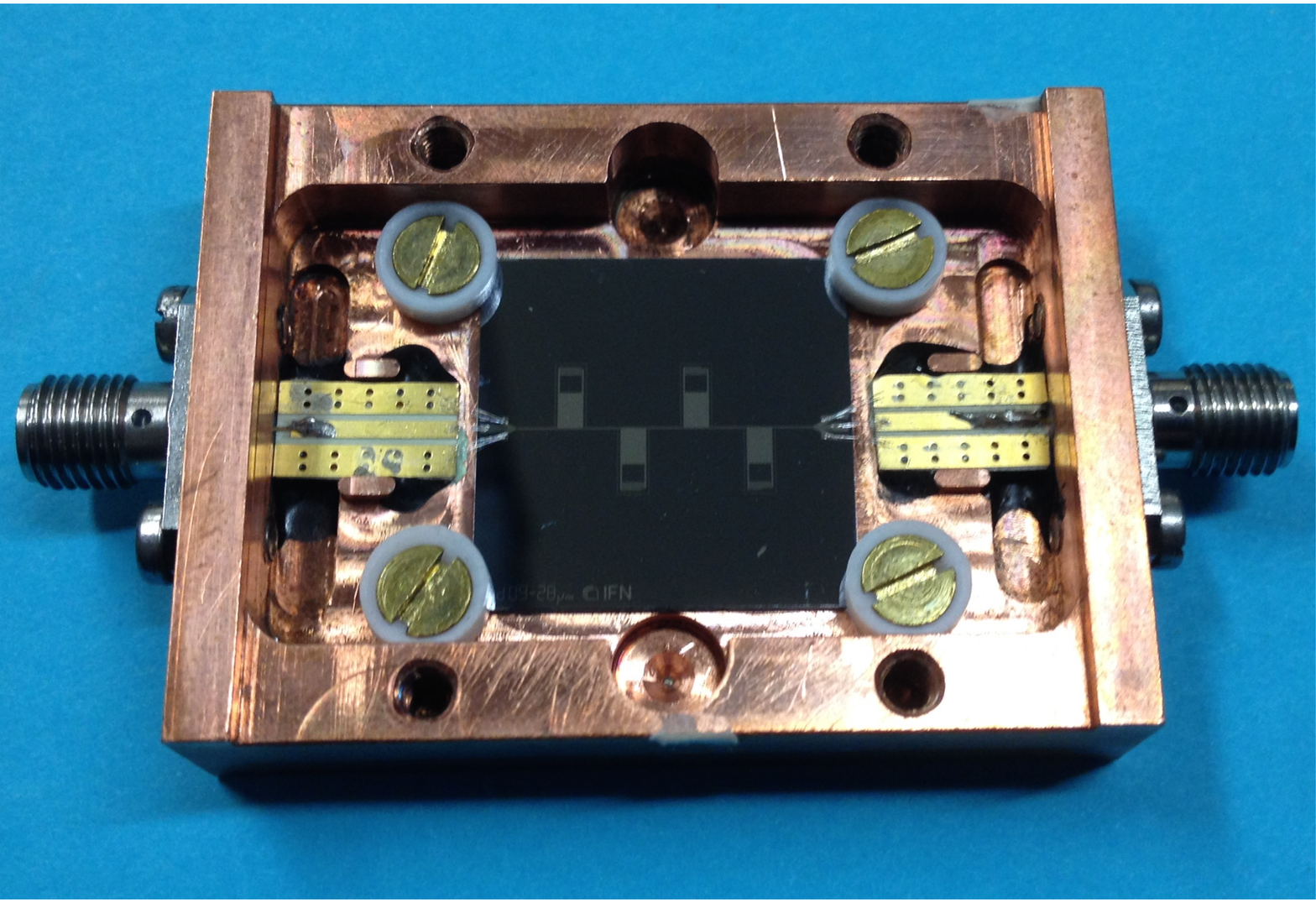}
 \caption{\label{fig:photo}The four Al KIDs deposited on the 2$\times$2\,cm$^2$ Si substrate. The chip is assembled in a copper structure using PTFE supports and illuminated from the back with a collimated \cobalt\ source and an optical fiber.}
\end{figure}

The detectors are fabricated at CNR IFN on high quality, \SubstrateThickness\ thick, high resistivity ($>$10\,k$\Omega\times$cm) Si(100) substrates.
The four lumped-element resonators (in the following \ka, \kb, \kc\ and \kd, ordered according to the position on the chip) are patterned by electron beam lithography
on a single 40\,nm thick Al film deposited using electron-gun evaporator.
The active area of the single pixel consists of an inductive meander made of 14 connected strips of 80\,$\mu$m$\times$2\,mm.
The meander is closed with a capacitor made of 5 interdigitated fingers of 1.2\,mm$\times$50\,$\mu$m, to ensure the uniformity of the current across the inductor.
The resonant frequency of each resonator is varied by cutting the last finger of the capacitor.
As shown in Figure~\ref{fig:photo}, the chip is assembled in a copper structure using PTFE supports with total contact area of about 3\,mm$^2$.
The other side of the holder (not shown) is covered with a copper collimator hosting a \cobalt\ calibration source (peaks at 6.4 and 14.4\,keV) and an optical fiber coupled to a room-temperature LED, that produces pulses at 400\,nm.
The source and the fiber are placed on the back of the substrate to avoid direct illumination of the resonators. The optical fiber points to the center of the substrate, while the \cobalt\ source is located nearby \kd.

The copper holder is thermally anchored to the coldest point of a $^3$He/$^4$He dilution refrigerator with base temperature of 10\,mK.

The output signal is fed into a CITLF4 SiGe cryogenic low noise amplifier\cite{amply} operated at 4\,K.
A detailed description of the chip design, the cryogenic setup of our laboratory at Sapienza University in Rome, the room-temperature electronics and the acquisition software can be found in references\cite{CalderWhitePaper,Bourrion:2011gi,Bourrion:2013ifa}.

\section{Data Analysis}
\label{sec:analysis}
\begin{figure}[htbp]
 \includegraphics[width=.45\textwidth, natwidth=500, natheight=345]{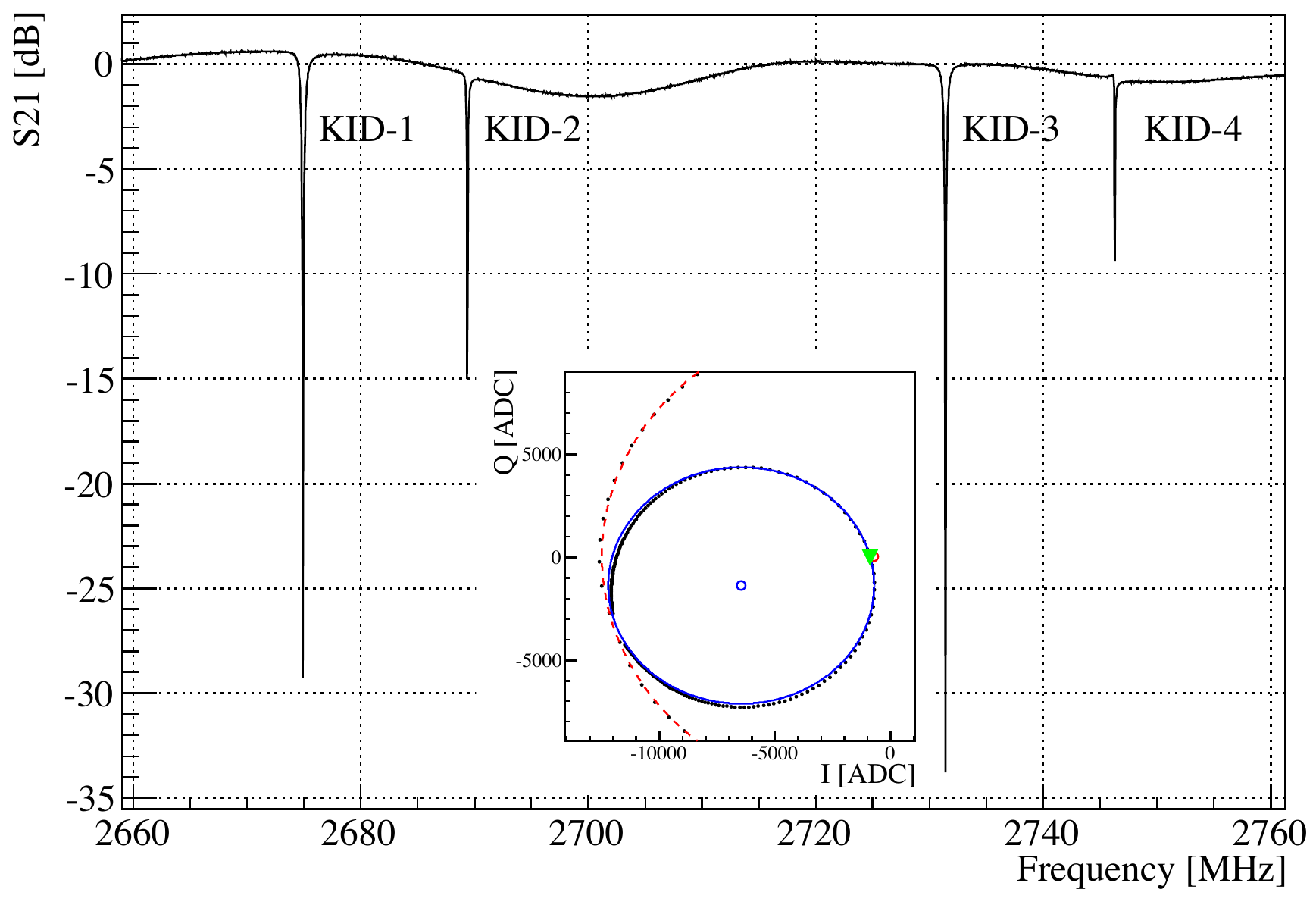} 
 \caption{\label{fig:resonances}Amplitude of the line transmission (S21) for a VNA frequency sweep around the resonances. Inset: fit of the resonance circle of \kc\ in the working point; the green marker indicates the resonant frequency.}
\end{figure}

A typical data collection consists in acquiring the complex transmission (S21) for a frequency sweep around the resonances (see Figure~\ref{fig:resonances}),
and fitting the resonance circles in order to extract the quality factor Q, the coupling quality factor Q$_c$ and the internal quality factor Q$_i$
using the method described in references~\cite{khalil2012,swenson2013}.

Resonators are designed to be over-coupled, thus the total quality factors Q (reported in Table~\ref{tab:QualityFactors}) are entirely dominated by the coupling quality factors Q$_c$.  
Q$_c$-values differ from the design value of 8000, likely due to the presence of parasitic slotline modes or standing waves between the detector and the cryogenic amplifier, or inter-resonator coupling resulting in coupled oscillation modes\cite{Noroozian2012}.
Q$_i$ is above 150$\times$10$^3$, but the values of Q$_c$ limit the accuracy of the estimation. 
\begin{table}[thb]
\caption{\label{tab:QualityFactors}Resonant frequency $f_0$, quality factor Q, optimal (off-resonance) microwave power at the amplifier input P$_{in}$ and \qp\ recombination time \tauqp. Errors on f$_0$ and P$_{in}$ are negligible, while errors on Q and \tauqp\ are dominated by fit systematics and are lower than 10$\%$.}
\begin{ruledtabular}
\begin{tabular}{lcccc}
                 &$f_0$	&Q              			&P$_{in}$		&\tauqp	\\					       							
		&[GHz]	&[$\times$10$^3$]		&[dBm]		&[$\mu$s] \\						              
\ka		&2.675	&6					&-63			&218	          \\	
\kb		&2.689	&18					&-64			&228	          \\
\kc		&2.731	&8					&-66			&243          \\		
\kd 		&2.746	&35					&-72			&233		  \\
\end{tabular}
\end{ruledtabular}
\end{table}

When the trigger of any of the resonators fires, we acquire a 2\,ms long time window for the real and imaginary parts of S21 for each resonator (I and Q) with a sampling frequency of 500\,kHz.
I and Q variations are then converted into changes in phase (\phase) and amplitude relative to the center of the resonance loop (blue marker in the circle reported in Figure~\ref{fig:resonances}).
In the following analysis we use only the \phase\ signal, as for this detector it is from 6 to 10 times larger than the amplitude one, depending on the KID.

To determine the optimal microwave power, we evaluate the signal-to-noise ratio scanning from -80\,dBm to -50\,dBm.
Increasing the input power produces a reduction of the noise contribution from the amplifier but, on the other hand,
decreases the \qp\ recombination time \tauqp\ and, as a consequence, the signal integration length.
The microwave power that optimizes the signal to noise ratio for each resonator is reported in Table~\ref{tab:QualityFactors}.

\begin{figure}[htbp]
 \includegraphics[width=.48\textwidth, natwidth=567, natheight=271]{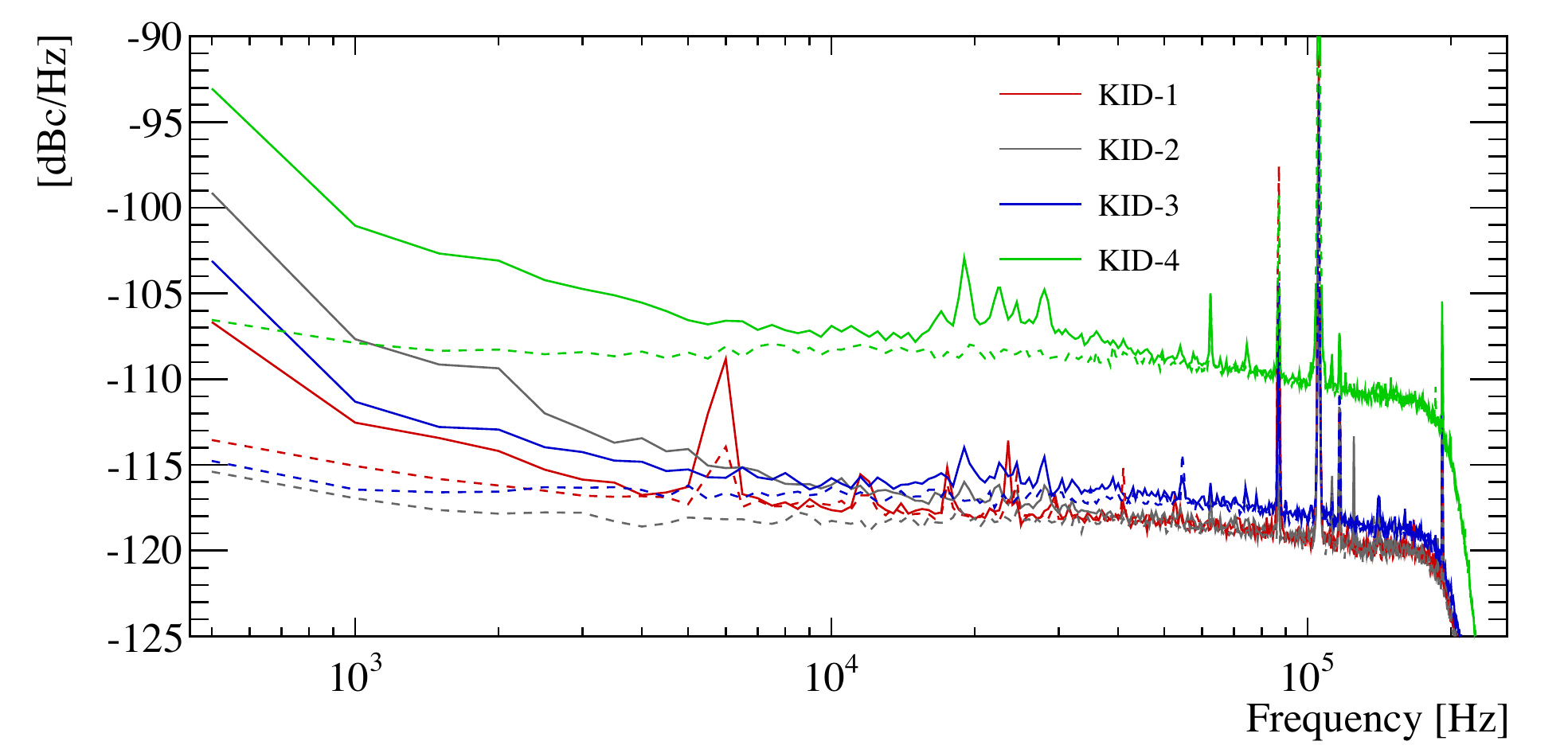}
 \caption{\label{fig:noise}Average noise power spectrum for phase (continuous line) and amplitude (dotted line) read-out. On top of the white noise from the amplifier, the phase noise exhibits an extra contribution at low frequency.}
\end{figure}

The average noise power spectrum is reported in Figure~\ref{fig:noise} for phase (continuous line) and amplitude (dotted line) read-out of each resonator.
The flat noise observed in the amplitude read-out and in the high frequency region of the phase read-out, 
is consistent with the noise temperature of the amplifier ($T_N\sim7\,K$).
The low-frequency region of the phase spectra is dominated by another noise source, whose origin is not clear yet.
It is not ascribable to two-level system noise, as it does not depend on temperature or microwave power. Furthermore, the presence of a mu-metal shield around the cryostat should guarantee an efficient suppression of noise due to static or low-frequency magnetic fields. Since a fraction of this noise is found to be correlated, it could be caused by frequency jitters in the read-out.

The high frequency noise in the acquired waveforms is rejected off-line using a software low-pass filter.
In order to avoid distortions in the rise-time of the pulses, the cut-off frequency is set at 100\,kHz ($\tau_{cut-off}\sim1.6$\,\musec).
Finally, the waveforms are processed with the optimum filter\cite{Gatti:1986cw,Radeka:1966}, which includes a resonator-specific rise time (see below). The results are not highly sensitive to the choice of rise time.

\begin{figure}[htbp]
 \includegraphics[width=.48\textwidth, natwidth=567, natheight=354]{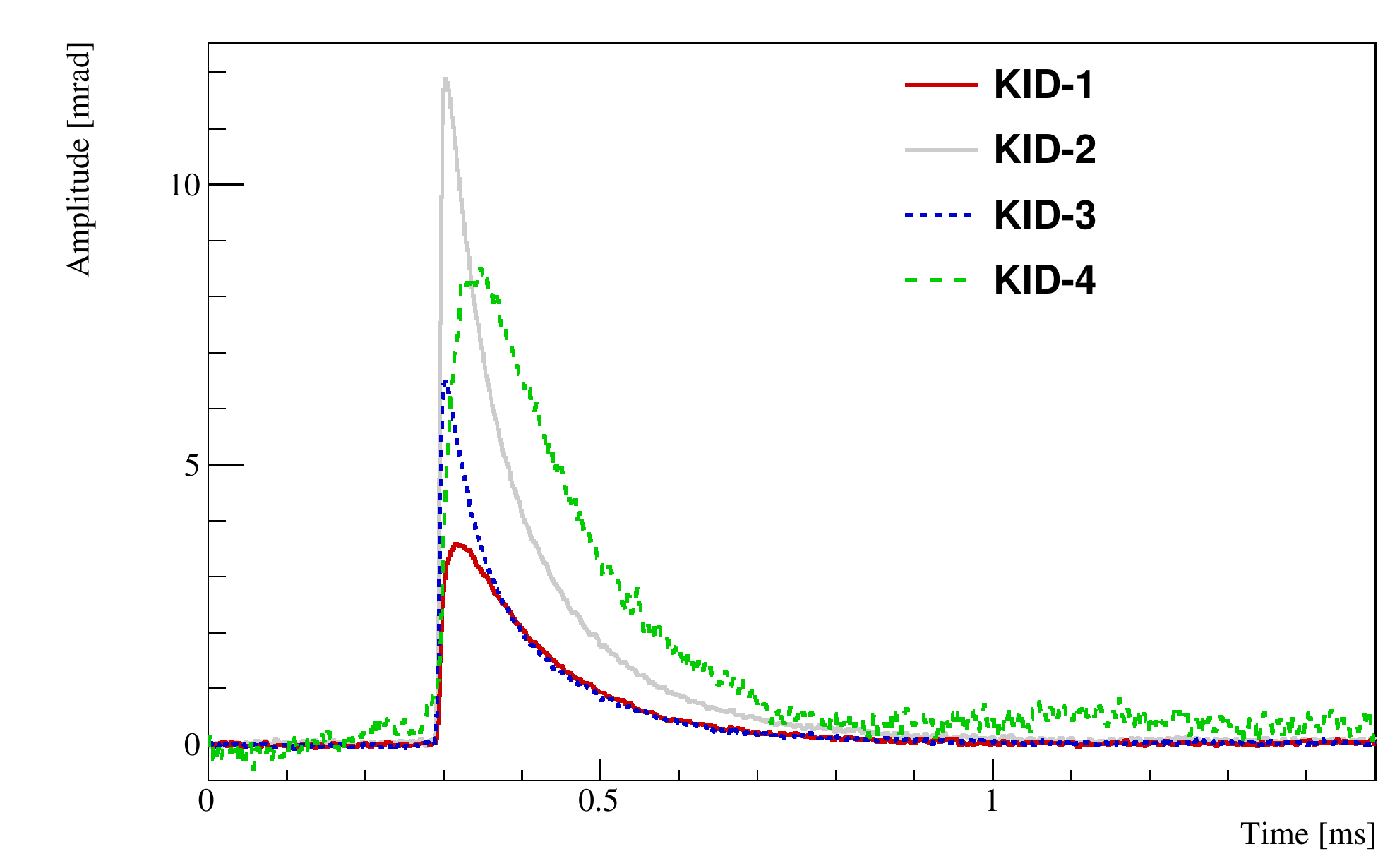}
 \caption{\label{fig:14keVpulse}Response of the four resonators to 15\,keV pulses produced by the optical fiber, placed in the proximity of \kb\ and \kc. The responses are obtained averaging many pulses to reduce the random noise.}
\end{figure}
In Figure~\ref{fig:14keVpulse} we show the typical response of the four resonators to the interaction of 15\,keV optical pulses, obtained by averaging several pulses to suppress the random noise contributions.

We fit the pulses with a model that includes the time constant of the low-pass filter, the ring-time of the resonator  ($\tau_r = Q/(\pi f_0)$) and two free parameters: a rise-time, which is related to the spread in the arrival time of phonons, and a decay time. 
As expected, the rise-time depends on the distance between the resonator and the optical fiber, and ranges from 2\,$\mu$s  (\kb\ and \kc) to 10\,$\mu$s (\ka) and 17\,$\mu$s (\kd). 
The decay time becomes faster increasing the microwave power or the temperature, and for this reason it is identified as \tauqp. 
This time constant does not depend on the energy of the optical pulses in the scan range (0.7-25\,keV), and its value is reported in Table~\ref{tab:QualityFactors} for each resonator.


\section{Energy calibration and efficiency}
\label{sec:results}
The energy $E$ absorbed in a resonator creates a number of \qp\ $\delta N_{qp} = \eta E/\Delta_0 $, where $\eta$ is the detection efficiency.
The variation $\delta N_{qp}$ produces a linear shift of the resonant frequency $f$ from the equilibrium one $f_0$\cite{Day:2003fk}: 
\begin{equation}
\label{eq:E}
E = \frac{\Delta_0}{\eta} \delta N_{qp} = \frac{ \Delta_0}{\eta} \left(  \frac{1}{p_0} \frac{f-f_0}{f_0}\right)
\end{equation}
where $p_0 = \alpha S_2(f,T)/4N_0 V\Delta_0$. 
The parameter $\alpha$ is the fraction of the total inductance due to kinetic inductance, $N_0$ is the single spin density of states (\Nzero) and $S_2(f,T)$ is a slow function of the temperature and of the resonant frequency that relates the phase variation of the complex conductivity to Cooper pairs breaking. 
In our working conditions, $S_2(f,T)$ is measured to be 2.3--2.6 depending on the resonator.
The active volume of the resonator $V$ is calculated by correcting the volume of the inductor (96500\,$\mu$m$^3$) for the average variation of current density (90\%) evaluated from SONNET simulations\cite{sonnet}.

We calculate $\Delta_0$ and $\alpha$ by measuring the variations of the resonant frequency $(f-f_0)/f_0$ as a function of the temperature, and fitting these data to the Mattis-Bardeen theory according to the procedure described by P\"{o}pel\cite{popel} and adopted by Gao \emph{et al.}\cite{GAOalpha}. The relationship between $(f-f_0)/f_0$ and $T$, indeed, depends only on $\Delta_0$ and $\alpha$, that are found to be the same for all the resonators. The average values are $\Delta_0=(201\pm6)\,\mu$eV and $\alpha=(5.8\pm0.5)\%$. 

We make a second, independent measurement of $\Delta_0$ by illuminating the chip with a Millimeter-Wave source.
We monitor phase, amplitude and resonant frequency of the resonators while increasing the frequency of the source from 75 to 110\,GHz.
The minimum source frequency that produces significant variations of the resonators features is $\nu_m = 95.5$\,GHz.
This value corresponds to $\Delta_0 = h\nu_m/2 =(197\pm5)\,\mu$eV, in full agreement with the previous value.
With these parameters, p$_0$ turns out to be $(1.2\pm0.1)\times10^{-13}$ for all the resonators.

Finally, equation~\ref{eq:E} can be modified in order to convert the frequency response $(f-f_0)/f_0$ into the measured phase variation $\delta\phi$ recalling that, for \phase\ measured in the center of the resonance loop, $(f-f_0)/f_0 = \delta\phi/4Q$:
\begin{equation}
\label{eq:AbsoluteCalib}
E =  \frac{\Delta_0}{4Q\eta p_0}\delta\phi.
\end{equation}
To estimate $\delta\phi$ we could use directly the amplitude of the phase signal.
However, to account for the observed small differences in the pulse shapes induced by the arrival time of phonons,
we evaluate $\delta\phi$ by integrating the time development of the pulse and dividing the result by \tauqp.

\begin{figure}[htbp]
    \includegraphics[width=.5\textwidth, natwidth=567, natheight=488]{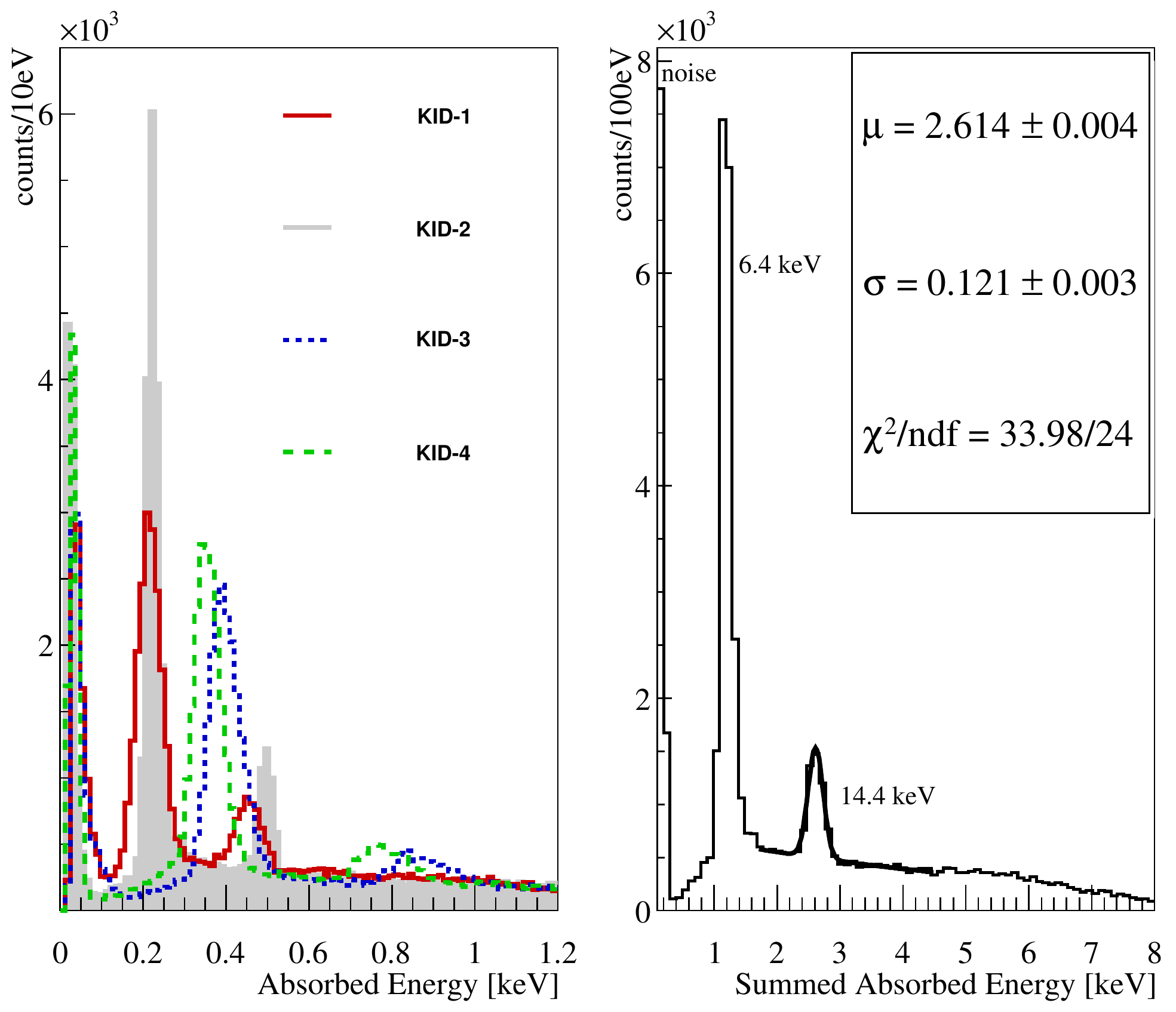}
    \caption{\label{fig:TotalEnergySpectrumAbsoluteCalib} Left: low energy spectrum of the energy absorbed in \ka\ (red, continuous line), \kb\ (gray, filled) \kc\ (blue, dotted line) and \kd\ (green, dashed line) calibrated with the function in Eq.~\ref{eq:AbsoluteCalib}; each resonator absorbs a different energy fraction of the X-rays from the \cobalt\ source depending on its position. Right: event by event sum of the the calibrated energies reported in the left plot; the fit of the 14.4\,keV peak produced by the \cobalt\ source is shown. The energy spectra are not corrected for the detection efficiency.}
\end{figure}

The energy-calibration function in Eq. \ref{eq:AbsoluteCalib} is applied to the single resonators to obtain the energy spectra of the \cobalt\  source (reported in Figure~\ref{fig:TotalEnergySpectrumAbsoluteCalib} (left)) and of the optical pulses. 
To infer the detection efficiency, we fit the position of the high energy peak produced by the \cobalt\  source, and we divide the result by the nominal energy of the line (see Table~\ref{tab:EnergyReso}). Applying this procedure to the lower energy peak (6.4\,keV) gives consistent results.
The estimation of the efficiency on  X-rays allows us to neglect uncertainties in the energy deposited, which could be significant when dealing with optical pulses.
To evaluate the total efficiency of the detector, we sum event-by-event the calibrated energies of each KID (see Figure~\ref{fig:TotalEnergySpectrumAbsoluteCalib} (right)) and fit the position of the 14.4\,keV peak, obtaining $\eta_{SUM}=2.6$\,keV/14.4\,keV = \efficiency. The error on the efficiency is dominated by the error on \tauqp\ introduced by the absolute calibration (Eq.~\ref{eq:AbsoluteCalib}).

We apply to optical pulses the analysis technique validated on X-rays.
The energy of the optical pulses was previously calibrated with a PMT at room temperature, and corrected for the PMT quantum efficiency, for the reflectivity of silicon\cite{SiIndex}, and for the geometrical efficiency, evaluated through a Monte Carlo simulation based on the Litrani software\cite{Latrina}.
We observe that the detector response is linear in the scan range.
The energies of the optical pulses evaluated using the absolute calibration (Eq.~\ref{eq:AbsoluteCalib}) and corrected for $\eta_{SUM}$, differ by less than 5$\%$ from what computed with the simulation.

In Table~\ref{tab:EnergyReso} we report the noise resolution $\sigma_{E}$, evaluated at the detector baseline, 
and the intrinsic noise of each resonator, calculated by scaling the resolution $\sigma_{E}$ for the single KID efficiency $\eta$.
Summing the response of the four KIDs on an event by event basis, we obtain a global energy resolution $\sigma_{E,SUM}$ of \NoiseGlobalResolution,
which corresponds to an intrinsic resolution of $\sigma_{E,SUM}\times\eta = 154\,eV\times 0.18 = 27.7$\,eV.
\begin{table}
\caption{\label{tab:EnergyReso}Measured noise resolution ($\sigma_{E}$), expected noise resolution in the limit in which the amplifier noise dominates ($\sigma_{E}^{amp}$) and integral of the optimally-filtered noise power spectral density ($\sigma_E^{PSD}$); pixel efficiency $\eta$ and intrinsic resolution of the resonator ($\sigma_E\times\eta$). In the last line we report the global performance of the detector.}
\begin{ruledtabular}
\begin{tabular}{lcccccc}
                 &$\sigma_E$	&$\sigma_{E}^{amp}$        &$\sigma_E^{PSD}$	&$\eta$			&$\sigma_E\times\eta$  \\
		&[eV]		 &[eV]	                          &[eV]				&[$\%$] 			&[eV]	                      \\
\ka		&400			&147		  	    		 &484				&3.1$\pm$0.4		&12.4    \\
\kb		&262			&60		             		 &253				&3.4$\pm$0.4 		&8.9        \\
\kc		&205			&75 		    	     	         &233				&6.1$\pm$0.7 		&12.5       \\
\kd 		&204			&50		             		 &184				&5.5$\pm$0.6		&11.2      \\
SUM 	&154			&42		            		 &					&18$\pm$2		&27.7        \\
\end{tabular}	
\end{ruledtabular}
\end{table}
This resolution is worse than the one expected if we were dominated by the amplifier noise that, for each resonator, can be computed as\cite{CalderWhitePaper}:
\begin{equation}
\sigma_{E}^{amp} = \frac{\Delta_0^2N_0V}{\eta\alpha Q S_2(f,T)} \sqrt{\frac{4Q_c^2}{Q^2} \frac{k_BT_N}{P_{in}\tau_{qp} } }\,.
\end{equation}
The discrepancy between the measured values $\sigma_E$ and $\sigma_{E}^{amp}$ can be interpreted as due to the excess low frequency noise in the phase readout.
This is supported by the fact that $\sigma_E$ matches closely $\sigma_E^{PSD}$, which is the integral of the optimally-filtered noise power spectral density.
Without this noise, we would have expected a global energy resolution of $\sigma_{E,SUM}^{amp} = \sqrt{\sum(\sigma_{E_i}^{amp}\times\eta_i)^2)}/\eta_{SUM} = 42\,\rm{eV}$.

\section{Conclusions}
In this paper we presented the results obtained with a 4-pixels Al array on Si substrate.
The comparison of the absolute calibration of the detector with the energy produced by a calibrated LED+optical system proved the reliability in reconstructing the energy of optical pulses from 0 to 25\,keV.
We derived an efficiency of \efficiency, that can be improved by increasing the active surface of the resonators.
The overall baseline resolution of \NoiseGlobalResolution\ is already competitive with some of the commonly used cryogenic light detectors, 
and can be further improved by suppressing the noise sources and, eventually, by using more sensitive superconductors.

\section*{Acknowledgements}
This work was supported by  the European Research Council (FP7/2007-2013) under contract  CALDER no. 335359
and  by the Italian Ministry of Research under the  FIRB  contract no. RBFR1269SL. We want to thank Dr. Silvio Morganti for his help in calibrating the optical system.

\bibliography{main}
\end{document}